\documentclass[pra,aps,amsmath,amssymb,twocolumn,showpacs,showkeys]{revtex4}
\usepackage{graphicx}
\newcommand{\bsg}{\boldsymbol{\sigma}}
\newcommand{\bro}{\boldsymbol{\rho}}
\newcommand{\pen}{\openone}

\newcommand{\wrh}{\widetilde{R}}
\newcommand{\wff}{\widetilde{F}}
\newcommand{\wth}{\widetilde{H}}

\begin{document}
\clearpage
\preprint{}

\title{Uncertainty and certainty relations for the Pauli observables in terms of the R\'{e}nyi entropies of order $\alpha\in(0;1]$}
\author{Alexey E. Rastegin}
\email{rast@api.isu.ru; alexrastegin@mail.ru}
\affiliation{Department of Theoretical Physics, Irkutsk State University,
Gagarin Bv. 20, Irkutsk 664003, Russia}

\begin{abstract}
We obtain uncertainty and certainty relations of state-independent
form for the three Pauli observables with use of the R\'{e}nyi
entropies of order $\alpha\in(0;1]$. It is shown that these
entropic bounds are tight in the sense that they are always
reached with certain pure states. A new result is the conditions
for equality in R\'{e}nyi-entropy uncertainty relations for the
Pauli observables. Upper entropic bounds in the pure-state case
are also novel. Combining the presented bounds leads to a band, in
which the rescaled average R\'{e}nyi $\alpha$-entropy ranges for a
pure measured state. A width of this band is compared with the
Tsallis formulation derived previously.
\end{abstract}

\pacs{03.67.-a, 03.65.Ta, 03.67.Hk}
\keywords{Pauli observables, R\'{e}nyi entropy, quantum measurement, uncertainty principle}

\maketitle

\section{Introduction}\label{sec1}

Heisenberg's uncertainty principle \cite{wh27} is one of the most
known results related to quantum incompatibility. Indeterminacy
relations are still the subject of active researchs
\cite{ozawa03,lahti}. Traditional relations pertain to
uncertainties of several observables in the same state. The
so-called state-extended uncertainty relations deal with one
observable and two different states \cite{chm12}. Due to the
papers \cite{deutsch,maass}, entropic functionals are widely used
in formulating the uncertainty principle \cite{ww10,brud11}. A
quantum state is generally characterized by the probabilities of
the outcomes of a test \cite{peresq}. In this sense, the entropic
formulation deals with quantum-mechanical primaries. Entropic
uncertainty relations in the presence of quantum memory are
formulated in Refs. \cite{BCCRR10,ccyz12,fan12}. Entropic lower
bounds of the papers \cite{rast105,rast12qic} pertain to a
situation substantial in quantum optics. Entropic trade-off
relations for a single quantum operation were examined in Refs.
\cite{rprz12,rast13a}. A majorization approach to entropic
uncertainty relations has been proposed \cite{fgg13,prz13}.

Entropic uncertainties for more than two observables characterize
a role of mutual unbiasedness \cite{ww10,aamb10}. Entropic
uncertainty bounds are essential in analyzing the security of
quantum cryptographic schemes \cite{dmg07,bfw11,ngbw12}.
Uncertainty relations for $(d+1)$ mutually unbiased bases in
$d$-dimensional Hilbert space were given in terms of the Shannon
entropy \cite{ivan92,jsan93,jsan95}. Entropic uncertainty
relations for mutually unbiased bases were considered
\cite{molm09}. The author of Ref. \cite{jsan93} derived exact
bounds for the qubit case $d=2$. Complementarity aspects of
entropic relations are considered in Refs. \cite{lars90,diaz}.
Generalizations of the Shannon entropy are also used in quantum
information theory \cite{bengtsson}. The R\'{e}nyi \cite{renyi61}
and Tsallis entropies \cite{tsallis} form important families of
one-parametric extensions. We have previously expressed
uncertainty and certainty relations for the Pauli observables in
terms of Tsallis' entropies \cite{rastqip13}. The writers of Ref.
\cite{zbp13} examined uncertainty relations for two qubit
observables in terms of Renyi's entropies with arbitrary orders.

In this work, we study lower and upper bounds on the sum of
R\'{e}nyi's $\alpha$-entropies, which quantify uncertainties in
measurement of the Pauli observables. The present work is further
development of the approach proposed in Refs.
\cite{rastijtp,rastqip13}. The paper is organized as follows. The
preliminary material is given in Sect. \ref{sec2}. In Sect.
\ref{sec3}, tight lower bounds on the sum of three R\'{e}nyi's
entropies of order $\alpha\in(0;1]$ are obtained. The conditions
for equality are developed as well. In Sect. \ref{sec4}, we
examine upper bounds on the sum of three $\alpha$-entropies in the
case of pure measured states. Tight upper bounds are derived for
all real $\alpha\in(0;1]$. Combining the lower and upper bounds
gives a band, in which the rescaled average R\'{e}nyi entropy
ranges in the pure-state case for $\alpha\in(0;1]$. In this
regard, we also compare the R\'{e}nyi and Tsallis formulations. In
Sect. \ref{sec5}, we conclude the paper with a summary of results.

\section{Preliminaries}\label{sec2}

In this section, the required material is reviewed. We will
quantify uncertainties of quantum measurements by means of the
R\'{e}nyi entropy. Let $p=\{p_{j}\}$ be a probability distribution
supported on $n$ points. For real $\alpha>0\neq1$, the R\'{e}nyi
$\alpha$-entropy is defined by \cite{renyi61}
\begin{equation}
R_{\alpha}(p):=\frac{1}{1-\alpha}{\ }{\ln}{\left(\sum\nolimits_{j=1}^{n} p_{j}^{\alpha}
\right)}
{\>}. \label{renent}
\end{equation}
It is a non-increasing function of order $\alpha$ \cite{renyi61}.
The Shannon entropy $H_{1}(p)=-\sum_{j}p_j\ln{p}_j$ is recovered
in the limit $\alpha\to1$. In the limit $\alpha\to\infty$, the
right-hand side of (\ref{renent}) leads to the min-entropy
\begin{equation}
R_{\infty}(p)=-\ln\bigl(\max{p}_{j}\bigr)
{\>}, \label{minren}
\end{equation}
where $\max{p}_{j}$ is the maximal probability. For
$\alpha\in(0;1)$, the right-hand side of (\ref{renent}) is a
concave function of probability distribution. Namely, for all
$\lambda\in[0;1]$ and two probability distributions $p=\{p_{j}\}$
and $q=\{q_{j}\}$, we have
\begin{equation}
R_{\alpha}\bigl(\lambda{p}+(1-\lambda)q\bigr)\geq
\lambda{\,}R_{\alpha}(p)+(1-\lambda)R_{\alpha}(q)
\ , \label{recn}
\end{equation}
whenever $0<\alpha<1$. We merely note that: (i) the function
$\xi\mapsto\xi^{\alpha}$ is concave for $\alpha\in(0;1)$, and (ii)
the function $\xi\mapsto(1-\alpha)^{-1}\ln\xi$ is increasing and
concave for $\alpha\in(0;1)$. Using Jensen's inequality twice,
these facts immediately leads to the property (\ref{recn}). The
standard case $\alpha=1$ also deals with the concave entropy. For
$\alpha>1$, however, the R\'{e}nyi $\alpha$-entropy is neither
purely convex nor purely concave \cite{ja04}. We also see that the
min-entropy (\ref{minren}) is convex, as the function
$\xi\mapsto-\ln\xi$. Applications of entropic measures in quantum
information theory are discussed in the book \cite{bengtsson}. The
standard entropies of partitions on quantum logic were considered
in Refs. \cite{yuan,zhma}. Within the R\'{e}nyi and Tsallis
formulations, this issue was developed in Ref. \cite{rastctp}.

Following \cite{rastijtp}, we further put the quantity
\begin{equation}
\Phi_{\alpha}(p):=\sum\nolimits_{j=1}^{n}p_j^{\alpha}
\ . \label{prerm}
\end{equation}
When $\alpha<\beta$, we have
$\Phi_{\alpha}(p)\geq\Phi_{\beta}(p)$. With respect to the
distribution $p=\{p_{j}\}$, the functional (\ref{prerm}) is
concave for $\alpha\in(0;1)$ and convex for
$\alpha\in(1;+\infty)$. The R\'{e}nyi entropy (\ref{renent}) can
be rewritten as
\begin{equation}
R_{\alpha}(p):=(1-\alpha)^{-1}\ln\Phi_{\alpha}(p)
\ . \label{renent1}
\end{equation}
The maximum $\ln{n}$ is reached with the equiprobable
distribution, when $p_{j}=1/n$ for all $1\leq{j}\leq{n}$. The
minimal zero value is reached with any deterministic distribution,
for which one of probabilities is $1$ and other are all zeros. We
will also use the Tsallis entropy. In terms of the functional
(\ref{prerm}), the Tsallis $\alpha$-entropy of degree
$\alpha>0\neq1$ is written as
\begin{equation}
H_{\alpha}(p)=\frac{\Phi_{\alpha}(p)-1}{1-\alpha}
\ . \label{taph}
\end{equation}
With the equiprobable distribution, the entropy (\ref{taph})
reaches its maximal value $\ln_{\alpha}(n)$. Here, the
$\alpha$-logarithm of positive $\xi$ is defined as
\begin{equation}
\ln_{\alpha}(\xi):=\frac{\xi^{1-\alpha}-1}{1-\alpha}
\ . \label{aldf}
\end{equation}
In the limit $\alpha\to1$, the $\alpha$-logarithm is reduced to
the usual logarithm. The Shannon entropy is obtained from
(\ref{taph}) also in this limit.

In the following, we will deal with the qubit case $n=2$. Here,
three complementary observables are usually represented by the
Pauli matrices $\bsg_{x}$, $\bsg_{y}$, $\bsg_{z}$, namely
\begin{equation}
\bsg_{x}=
\begin{pmatrix}
0 & 1 \\
1 & 0
\end{pmatrix}
{\,}, \quad
\bsg_{y}=
\begin{pmatrix}
0 & -i \\
i & 0
\end{pmatrix}
{\,}, \quad
\bsg_{z}=
\begin{pmatrix}
1 & 0 \\
0 & -1
\end{pmatrix}
{\,}. \label{pmtr}
\end{equation}
These matrices traditionally used for describing spin-$1/2$
observables. Each of the matrices has the eigenvalues $\pm1$. By
$\bigl\{|0\rangle,|1\rangle\bigr\}$, we mean the eigenbasis of
$\bsg_{z}$, that is
\begin{equation}
|0\rangle=
\begin{pmatrix}
1  \\
0
\end{pmatrix}
{\,}, \qquad
|1\rangle=
\begin{pmatrix}
0  \\
1
\end{pmatrix}
{\,}. \label{01bs}
\end{equation}
The eigenstates of $\bsg_{x}$ and $\bsg_{y}$ are
written as
\begin{equation}
|x_{\pm}\rangle=\frac{1}{\sqrt{2}}
\begin{pmatrix}
1  \\
\pm1
\end{pmatrix}
{\,}, \qquad
|y_{\pm}\rangle=\frac{1}{\sqrt{2}}
\begin{pmatrix}
1  \\
\pm{i}
\end{pmatrix}
{\,}. \label{xybb}
\end{equation}
Here, we have $\bsg_{x}|x_{\pm}\rangle=\pm|x_{\pm}\rangle$ and
$\bsg_{y}|y_{\pm}\rangle=\pm|y_{\pm}\rangle$. The three bases
given by (\ref{01bs}) and (\ref{xybb}) are mutually unbiased.
Measurements in these eigenbases are used in six-state
cryptographic protocols \cite{ngbw12,bfw11}.

We now write the probabilities corresponding to measurement of
each of the observables $\bsg_{x}$, $\bsg_{y}$, $\bsg_{z}$. Up to
a unimodular factor, we can represent a normalized pure state in
the form
\begin{equation}
|\psi\rangle=\cos\tau|0\rangle+e^{i\varphi}\sin\tau|1\rangle=
\begin{pmatrix}
\cos\tau  \\
e^{i\varphi}\sin\tau
\end{pmatrix}
{\,}, \label{pswr}
\end{equation}
where $\tau$ and $\varphi$ are real numbers. Assuming
$\varphi\in[0;2\pi)$, we will take $\tau\in[0;\pi/2]$, since a
global phase in the state vector has no physical relevance. For
the observables $\bsg_{x}$ and $\bsg_{y}$, the probabilities are
respectively given as $|\langle{x}_{\pm}|\psi\rangle|^{2}$ and
$|\langle{y}_{\pm}|\psi\rangle|^{2}$. The final expressions are
obtained in the form \cite{rastqip13}
\begin{align}
p_{\pm}&=\frac{1\pm\sin2\tau\cos\varphi}{2}
\ , \label{xpr}\\
q_{\pm}&=\frac{1\pm\sin2\tau\sin\varphi}{2}
\ , \label{ypr}\\
r_{\pm}&=\frac{1\pm\cos2\tau}{2}
\ . \label{zpr}
\end{align}
Substituting (\ref{xpr}), (\ref{ypr}), (\ref{zpr}) into the
right-hand side of (\ref{renent}), one gives the three entropies
$R_{\alpha}\bigl(\bsg_{x}|\psi\bigr)$,
$R_{\alpha}\bigl(\bsg_{y}|\psi\bigr)$,
$R_{\alpha}\bigl(\bsg_{z}|\psi\bigr)$ for the state (\ref{pswr}).
We will study lower and upper bounds on the sum of such entropies
for $\alpha\in(0;1]$.

\section{Tight lower bounds on the sum of entropies of degree $\alpha\in(0;1]$}\label{sec3}

In this section, we derive tight lower bounds on the sum of three
R\'{e}nyi entropies of order $\alpha\in(0;1]$. A desired bound
will firstly be obtained for pure states of the form (\ref{pswr}),
when the probabilities are given by (\ref{xpr}), (\ref{ypr}), and
(\ref{zpr}). Using the concavity properties, we then extend the
result to all mixed states of a qubit. For $\alpha>0\neq1$, we
introduce the function
\begin{equation}
F_{\alpha}(\tau,\varphi)=
\Phi_{\alpha}(p){\,}\Phi_{\alpha}(q){\,}\Phi_{\alpha}(r)
\ , \label{fnpd}
\end{equation}
in which we substitute (\ref{xpr}), (\ref{ypr}), and (\ref{zpr}).
Using this function, the entropic sum is rewritten as
\begin{equation}
R_{\alpha}\bigl(\bsg_{x}|\psi\bigr)+R_{\alpha}\bigl(\bsg_{y}|\psi\bigr)
+R_{\alpha}\bigl(\bsg_{z}|\psi\bigr)
=\frac{1}{1-\alpha}{\>}\ln{F}_{\alpha}(\tau,\varphi)
{\>}. \label{fnpd1}
\end{equation}
Since the function $\xi\mapsto(1-\alpha)^{-1}\ln\xi$ increases for
$\alpha\in(0;1)$, we aim to minimize (\ref{fnpd}) in the domain of
interest. As was already noted, the variables are initially in the
intervals $\tau\in[0;\pi/2]$ and $\varphi\in[0;2\pi)$. In the task
of optimization, however, we can restrict a consideration to the
rectangular domain \cite{rastqip13}
\begin{equation}
D:=\bigl\{
(\tau,\varphi):{\>}\tau\in[0;\pi/4],{\>}\varphi\in[0;\pi/4]
\bigr\}
\ . \label{dotv}
\end{equation}
Here, we claim that in the total domain
$\bigl\{(\tau,\varphi):{\>}\tau\in[0;\pi/2],{\>}\varphi\in[0;2\pi)\bigr\}$,
the function  (\ref{fnpd}) takes the same range of values as in
the domain (\ref{dotv}). The reasons are the following. Taking
$\tau\in[0;\pi/2]$ and $\varphi\in(\pi;2\pi)$, one first uses
$\varphi\mapsto\varphi-\pi$. The latter merely swaps two values in
each of the pairs (\ref{xpr}) and (\ref{ypr}), without altering
$\Phi_{\alpha}(p)$ and $\Phi_{\alpha}(q)$. So, we restrict to
$\varphi\in[0;\pi]$. Taking further $\tau\in[0;\pi/2]$ and
$\varphi\in(\pi/2;\pi]$, one applies $\varphi\mapsto\pi-\varphi$.
Then the probabilities $p_{\pm}$ are swapped and the probabilities
$q_{\pm}$ are the same, with no changes in (\ref{fnpd}). Hence, we
restrict to $\varphi\in[0;\pi/2]$. Acting in
$\varphi\in(\pi/4;\pi/2]$, mapping $\varphi\mapsto\pi/2-\varphi$
implies swapping $p_{j}$ and $q_{j}$ for $j=\pm$, without altering
the product $\Phi_{\alpha}(p){\,}\Phi_{\alpha}(q)$. So, we can
focus on $\tau\in[0;\pi/2]$ and $\varphi\in[0;\pi/4]$. Finally,
mapping $\tau\mapsto\pi/2-\tau$ for $\tau\in(\pi/4;\pi/2]$ does
not alter $\sin2\tau$ and reverses the sign of $\cos2\tau$. So,
each of the three multipliers in the right-hand side of
(\ref{fnpd}) remains unchanged. The following statement takes
place.

\newtheorem{thm31}{Theorem}
\begin{thm31}\label{inra01}
Let qubit state be described by density matrix $\bro$. For all
$\alpha\in(0;1]$, the entropic sum satisfies
\begin{equation}
R_{\alpha}\bigl(\bsg_{x}|\bro\bigr)+R_{\alpha}\bigl(\bsg_{y}|\bro\bigr)+
R_{\alpha}\bigl(\bsg_{z}|\bro\bigr)\geq2\ln2
\ , \label{ral0}
\end{equation}
with equality if and only if the qubit state is an eigenstate of
either of the observables $\bsg_{x}$, $\bsg_{y}$, $\bsg_{z}$.
\end{thm31}

{\bf Proof.} Since the case of the Shannon entropy was already
given in theorem 1 of the paper \cite{rastqip13}, we further
assume $\alpha\in(0;1)$. We will show that the right-hand side of
(\ref{ral0}) is equal to the minimum of (\ref{fnpd1}) in the domain
(\ref{dotv}). Differentiating (\ref{fnpd}) with respect to
$\varphi$, we have
\begin{align}
&\frac{\partial}{\partial\varphi}{\>}F_{\alpha}(\tau,\varphi)=
\nonumber\\
&F_{\alpha}(\tau,\varphi)\left(
\frac{1}{\Phi_{\alpha}(p)}{\>}
\frac{\partial\Phi_{\alpha}(p)}{\partial\varphi}
+\frac{1}{\Phi_{\alpha}(q)}{\>}
\frac{\partial\Phi_{\alpha}(q)}{\partial\varphi}
\right)
{\,}. \label{drfa}
\end{align}
Usual calculations show that
\begin{align}
\frac{1}{\Phi_{\alpha}(p)}{\>}
\frac{\partial\Phi_{\alpha}(p)}{\partial\varphi}&=
-\frac{\alpha}{2}{\>}
\frac{p_{+}^{\alpha-1}-p_{-}^{\alpha-1}}{p_{+}^{\alpha}+p_{-}^{\alpha}}
{\>}\sin2\tau\sin\varphi
\ , \label{drpp}\\
\frac{1}{\Phi_{\alpha}(q)}{\>}
\frac{\partial\Phi_{\alpha}(q)}{\partial\varphi}&=
+\frac{\alpha}{2}{\>}
\frac{q_{+}^{\alpha-1}-q_{-}^{\alpha-1}}{q_{+}^{\alpha}+q_{-}^{\alpha}}
{\>}\sin2\tau\cos\varphi
\ . \label{drqq}
\end{align}
Introducing the variables $u=\sin2\tau\cos\varphi$ and
$v=\sin2\tau\sin\varphi$, we rewrite (\ref{drfa}) as
\begin{equation}
\frac{1}{F_{\alpha}(\tau,\varphi)}{\>}
\frac{\partial}{\partial\varphi}{\>}F_{\alpha}(\tau,\varphi)=
\alpha{uv}\left(
\frac{f_{\alpha}(u)}{g_{\alpha}(u)}-\frac{f_{\alpha}(v)}{g_{\alpha}(v)}
\right)
{\,}. \label{drfa1}
\end{equation}
Here, we use the functions
\begin{align}
f_{\alpha}(u)&=u^{-1}\left((1-u)^{\alpha-1}-(1+u)^{\alpha-1}\right)
{\>}, \label{fudf}\\
g_{\alpha}(u)&=(1+u)^{\alpha}+(1-u)^{\alpha}
{\>}. \label{gudf}
\end{align}
For $\alpha\in(0;1)$, the function $f_{\alpha}(u)$ monotonically
increases, whereas the function $g_{\alpha}(u)$ monotonically
decreases with $u\in[0;1]$. To prove the claim, these functions
are expanded into power series about the origin. Using the
binomial theorem and properties of the binomial coefficients, we
have
\begin{equation}
f_{\alpha}(u)=2(1-\alpha)+2\sum_{k=1}^{\infty}\binom{2k+1-\alpha}{2k+1} u^{2k}
\ . \label{al01}
\end{equation}
We stress that this series contains only strictly positive
coefficients. Indeed, for $k\geq1$ and $\alpha\in(0;1)$ we have
\begin{equation}
\binom{2k+1-\alpha}{2k+1}=\frac{(2k+1-\alpha)\cdots(2-\alpha)(1-\alpha)}{(2k+1)!}>0
\ . \label{bi01}
\end{equation}
Hence, the function (\ref{al01}) monotonically increases. Further,
we obtain the expansion
\begin{equation}
g_{\alpha}(u)=2-2\sum_{k=1}^{\infty} c_{2k}{\,}u^{2k}
\ . \label{al02}
\end{equation}
For $k\geq1$ and $\alpha\in(0;1)$, the coefficients $c_{2k}$ are
strictly positive, i.e.
\begin{align}
c_{2k}&=(-1)\binom{\alpha}{2k}
\nonumber\\
&=\alpha{\,}\frac{(2k-1-\alpha)\cdots(2-\alpha)(1-\alpha)}{(2k)!}>0
\ . \label{bi02}
\end{align}
So, the function (\ref{al02}) monotonically decreases. Hence, the
ratio $f_{\alpha}(u)/g_{\alpha}(u)$ is monotonically increasing
function of $u\in[0;1]$. The inequality $v<u$ then gives that the
right-hand side of (\ref{drfa1}) is strictly positive. In the
interior of the domain (\ref{dotv}), the function
$F_{\alpha}(\tau,\varphi)$ increases with $\varphi$. On the
boundary lines $\tau=0$ and $\tau=\pi/4$, we respectively have
$\partial{F}_{\alpha}/\partial\varphi=0$ and
$\partial{F}_{\alpha}/\partial\varphi\geq0$. These points implies
that the minimal and maximal values of $F_{\alpha}(\tau,\varphi)$
in the domain (\ref{dotv}) are reached on the lines $\varphi=0$
and $\varphi=\pi/4$, respectively.

To find the minimum, we substitute $\varphi=0$ and obtain
probabilities
\begin{equation}
p_{\pm}=\frac{1\pm\sin2\tau}{2}
\ , \qquad
q_{\pm}=\frac{1}{2}
\ , \qquad
r_{\pm}=\frac{1\pm\cos2\tau}{2}
\ ,  \label{prvp0}
\end{equation}
whence $\Phi_{\alpha}(q)=2^{1-\alpha}$. Differentiating with
respect to $\tau$, we further write
\begin{align}
&\frac{\partial}{\partial\tau}{\>}F_{\alpha}(\tau,0)=
\nonumber\\
&F_{\alpha}(\tau,0)\left(
\frac{1}{\Phi_{\alpha}(p)}{\>}
\frac{\partial\Phi_{\alpha}(p)}{\partial\tau}
+\frac{1}{\Phi_{\alpha}(r)}{\>}
\frac{\partial\Phi_{\alpha}(r)}{\partial\tau}
\right)
{\,}. \label{drfb}
\end{align}
Using (\ref{prvp0}), we easily obtain
\begin{align}
\frac{1}{\Phi_{\alpha}(p)}{\>}
\frac{\partial\Phi_{\alpha}(p)}{\partial\tau}&=
+\alpha{\,}
\frac{p_{+}^{\alpha-1}-p_{-}^{\alpha-1}}{p_{+}^{\alpha}+p_{-}^{\alpha}}
{\>}\cos2\tau
\ , \label{drpb}\\
\frac{1}{\Phi_{\alpha}(r)}{\>}
\frac{\partial\Phi_{\alpha}(r)}{\partial\tau}&=
-\alpha{\,}
\frac{r_{+}^{\alpha-1}-r_{-}^{\alpha-1}}{r_{+}^{\alpha}+r_{-}^{\alpha}}
{\>}\sin2\tau
\ . \label{drrb}
\end{align}
Denoting $u=\cos2\tau$ and $v=\sin2\tau$, we rewrite (\ref{drfb})
as
\begin{equation}
\frac{1}{F_{\alpha}(\tau,0)}{\>}
\frac{\partial}{\partial\tau}{\>}F_{\alpha}(\tau,0)=
2\alpha{uv}\left(\frac{f_{\alpha}(u)}{g_{\alpha}(u)}
-\frac{f_{\alpha}(v)}{g_{\alpha}(v)}
\right)
{\,}. \label{drfb1}
\end{equation}
As $u>v$ for $\tau\in(0;\pi/8)$ and $u<v$ for
$\tau\in(\pi/8;\pi/4)$, the derivative (\ref{drfb1}) is strictly
positive in the former interval and strictly negative in the
latter one. So, the minimal value of $F_{\alpha}(\tau,0)$ is
reached at the end points of the interval $\tau\in[0;\pi/4]$. In
both the points, the function (\ref{fnpd}) is equal to
\begin{equation}
F_{\alpha}(0,0)=F_{\alpha}(\pi/4,0)=2^{2(1-\alpha)}
\ . \label{f040}
\end{equation}
Combining this with (\ref{fnpd1}) immediately leads to the
inequality (\ref{ral0}) for all pure states. This bound remains
valid for mixed states due to concavity of the R\'{e}nyi entropy
of order $\alpha\in(0;1)$.

Let us prove conditions for equality. We first prove the claim for
the case of pure measured state (\ref{pswr}). In the domain
(\ref{dotv}), the function $F_{\alpha}(\tau,\varphi)$ takes its
minimum (\ref{f040}) only at the points $\tau=\varphi=0$ and
$\tau=\pi/4$, $\varphi=0$. In both the points, one of the
distributions $\{p_{\pm}\}$, $\{q_{\pm}\}$, $\{r_{\pm}\}$ is
deterministic and other two are equiprobable. This is the only
situation, in which the minimum of $F_{\alpha}(\tau,\varphi)$
holds. It is seen from (\ref{prvp0}) that the distribution
$\{q_{\pm}\}$ is equiprobable for the above two points. The total
domain
$\bigl\{(\tau,\varphi):{\>}\tau\in[0;\pi/2],{\>}\varphi\in[0;2\pi)\bigr\}$
for the state (\ref{pswr}) contains also points, in which the
distribution $\{q_{\pm}\}$ is deterministic and other two are
equiprobable. In any case, the minimum is reached only if one of
the three distributions is deterministic. Clearly, this condition
is sufficient as well. To saturate the inequality (\ref{ral0}),
the state $|\psi\rangle$ should be an eigenstate of one of the
observables $\bsg_{x}$, $\bsg_{y}$, $\bsg_{z}$.

We will further prove that the inequality (\ref{ral0}) cannot be
saturated with impure states. Let the spectral decomposition of
impure $\bro$ be written as
\begin{equation}
\bro=\lambda_{+}|\psi_{+}\rangle\langle\psi_{+}|+\lambda_{-}|\psi_{-}\rangle\langle\psi_{-}|
\ . \label{spdc}
\end{equation}
Here, the eigenstates are mutually orthogonal and the strictly
positive eigenvalues obey the condition
$\lambda_{+}+\lambda_{-}=1$. Since the entropy (\ref{renent}) is
concave for $\alpha\in(0;1)$, we obtain
\begin{align}
&\sum_{\nu=x,y,z}R_{\alpha}\bigl(\bsg_{\nu}|\bro\bigr)\geq
\nonumber\\
&\lambda_{+}\sum_{\nu=x,y,z}R_{\alpha}\bigl(\bsg_{\nu}|\psi_{+}\bigr)+
\lambda_{-}\sum_{\nu=x,y,z}R_{\alpha}\bigl(\bsg_{\nu}|\psi_{-}\bigr)
\ . \label{nxyz}
\end{align}
If for any of the states $|\psi_{\pm}\rangle$ the entropic sum
does not reach the lower bound $2\ln{2}$, the left-hand side of
(\ref{nxyz}) does not reach this bound as well. Hence, the
question is quite reduced to the case, when the matrix $\bro$ is
diagonal with respect to eigenbasis of either of the $\bsg_{x}$,
$\bsg_{y}$, $\bsg_{z}$. For definiteness, we assume that
$|\psi_{\pm}\rangle=|x_{\pm}\rangle$. Measuring each of the
observables $\bsg_{y}$ and $\bsg_{z}$ then results in the
equiprobable distribution, whence
$R_{\alpha}\bigl(\bsg_{y}|\bro\bigr)=R_{\alpha}\bigl(\bsg_{z}|\bro\bigr)=\ln{2}$.
For the measurement of $\bsg_{x}$, we have outcomes $\pm1$ with
probabilities $\lambda_{\pm}$, respectively. For $\alpha\in(0;1)$,
the first derivative of the function
$x\mapsto{x}^{\alpha}+(1-x)^{\alpha}$ is strictly positive for
$0<x<1/2$ and strictly negative for $1/2<x<1$. Except for
$\lambda_{+}=0$ and $\lambda_{+}=1$, we then have
$\lambda_{+}^{\alpha}+\lambda_{-}^{\alpha}=\Phi_{\alpha}(\lambda)>1$
and $R_{\alpha}\bigl(\bsg_{x}|\bro\bigr)>0$. The latter implies
that the sum of three entropies is strictly larger than $2\ln{2}$.
$\blacksquare$

Theorem \ref{inra01} provides a lower bound on the sum of three
R\'{e}nyi's entropies for all $\alpha\in(0;1]$. This bound is
tight in the sense that it is certainly reached with an eigenstate
of one of the Pauli observables. Previously, the standard case
$\alpha=1$ has been considered in \cite{jsan93} and, as a
particular case of the Tsallis formulation, also in
\cite{rastqip13}. Namely, we have the lower bound
\begin{equation}
H_{1}\bigl(\bsg_{x}|\bro\bigr)+H_{1}\bigl(\bsg_{y}|\bro\bigr)+
H_{1}\bigl(\bsg_{z}|\bro\bigr)\geq2\ln2
\ . \label{shal0}
\end{equation}
For $\alpha\in(0;1)$, the inequality (\ref{ral0}) could be derived
from (\ref{shal0}) due to the fact that the Renyi $\alpha$-entropy
does not increase with $\alpha$. In this way, however, we cannot
resolve conditions for equality. The above method allows to
formulate such conditions. Thus, we obtained tight uncertainty
relations for the Pauli observables in terms of R\'{e}nyi's
entropies of order $\alpha\in(0;1]$. A utility of entropic bounds
with a parametric dependence was noted in \cite{maass}. In
particular, this dependence allows to find more exactly the domain
of acceptable values for unknown probabilities with respect to
known ones. Some studies were devoted to uncertainty relations of
state-dependent form. For example, the writers of \cite{clz11}
have derived a stronger bound, in which the right-hand side of
(\ref{shal0}) is added by the von Neumann entropy of $\bro$. For
mutually unbiased bases, state-dependent uncertainty relations in
terms of the Shannon entropies were derived in \cite{molm09}. Such
uncertainty relations have been extended to both the R\'{e}nyi and
Tsallis formulations \cite{rast13b}. A dependence of lower
entropic bounds on a degree of impurity of the measured density
matrix deserves further investigations.

\section{Tight upper bounds in the pure-state case for $\alpha\in(0;1]$}\label{sec4}

In this section, we will study upper bounds on the sum of three
R\'{e}nyi's entropies for the Pauli observables. Certainly, such
bounds essentially depend on a type of considered states. The
completely mixed state is described by density operator
$\bro_{*}=\pen/2$, where $\pen$ is the identity $2\times2$-matrix.
Measuring each of the observables $\bsg_{x}$, $\bsg_{y}$,
$\bsg_{z}$ in this state will lead to the equiprobable
distribution. With this distribution, the entropy (\ref{renent})
takes its maximal possible value $\ln{2}$. For all $\alpha>0$ and
arbitrary $\bro$, therefore, we have the upper bound
\begin{equation}
\sum_{\nu=x,y,z}R_{\alpha}\bigl(\bsg_{\nu}|\bro\bigr)\leq
\sum_{\nu=x,y,z}R_{\alpha}\bigl(\bsg_{\nu}|\bro_{*}\bigr)
=3\ln{2}
\ . \label{rscm}
\end{equation}
Further, we ask for upper entropic bounds in the case of pure
states. Modifying the method of previous section, one will obtain
tight bounds from above for real $\alpha\in(0;1]$. Following
\cite{rastqip13}, we recall an intuitive reason that makes the
result physically reasonable. Mathematically, we aim to maximize
the function (\ref{fnpd1}) in the domain (\ref{dotv}). According to
the proof of Theorem \ref{inra01}, the maximum is reached on the
line $\varphi=\pi/4$. Substituting the latter into the formulas
(\ref{xpr}), (\ref{ypr}), and (\ref{zpr}), we obtain the
probabilities
\begin{equation}
p_{\pm}=q_{\pm}=\frac{1\pm{v}}{2}
\ , \qquad
r_{\pm}=\frac{1\pm{u}}{2}
\ ,  \label{prvp2}
\end{equation}
where $u=\cos2\tau$, $v=\sin2\tau/\sqrt{2}$. Here, the variables
$u$ and $v$ satisfy the condition
\begin{equation}
u^{2}+2v^{2}=1
\ . \label{uvcd}
\end{equation}
Due to (\ref{prvp2}), the distributions $\{p_{\pm}\}$ and
$\{q_{\pm}\}$ should concur for maximizing the entropic sum in the
case of pure states and considered values of $\alpha$. For impure
states, the maximum (\ref{rscm}) is reached only if the
probability distributions are all equiprobable and herewith
identical. In the case of distributions (\ref{prvp2}), we can
assume the following \cite{rastqip13}. The maximum takes place,
when the distribution $\{r_{\pm}\}$ also concurs with
$\{p_{\pm}\}=\{q_{\pm}\}$, i.e. $u=v$. Combining the latter with
(\ref{uvcd}) gives $u=v=1/\sqrt{3}$. The R\'{e}nyi
$\alpha$-entropy of each of three probability distributions is
equal to
\begin{equation}
\wrh_{\alpha}=\frac{1}{1-\alpha}{\ }{\ln}{\left\{
\left(\frac{1+1/\sqrt{3}}{2}\right)^{\alpha}+
\left(\frac{1-1/\sqrt{3}}{2}\right)^{\alpha}
\right\}}
{\ }. \label{prore}
\end{equation}
This hint is fruitful in the case of Tsallis' entropies
\cite{rastqip13}. It can be adapted for the R\'{e}nyi case as
well.

\newtheorem{thm41}[thm31]{Theorem}
\begin{thm41}\label{upral}
Let qubit state be described by the ket $|\psi\rangle$. For all
$\alpha\in(0;1]$, the entropic sum obeys
\begin{equation}
R_{\alpha}\bigl(\bsg_{x}|\psi\bigr)+R_{\alpha}\bigl(\bsg_{y}|\psi\bigr)
+R_{\alpha}\bigl(\bsg_{z}|\psi\bigr)
\leq3{\,}\wrh_{\alpha}
\ . \label{urwh}
\end{equation}
Equality holds if and only if the three probability distributions
are all, up to swapping, the pair $\left(1\pm1/\sqrt{3}\right)/2$.
\end{thm41}

{\bf{Proof.}} We again assume $\alpha\in(0;1)$, since the case of
the Shannon entropy was fully analyzed in theorem 5 of
\cite{rastqip13}. Using the probabilities (\ref{prvp2}), we
rewrite the product (\ref{fnpd}) in the form
\begin{equation}
\wff_{\alpha}(u,v)=2^{-3\alpha}
g_{\alpha}(u){\,}g_{\alpha}(v)^{2}
{\>}. \label{fnpd2}
\end{equation}
When $\tau\in[0;\pi/4]$, the variables $u$ and $v$ lie in the
interval $[0;1]$. The function (\ref{fnpd2}) should be maximized
in this interval under the condition (\ref{uvcd}). By
(\ref{uvcd}), we have that $du/dv=-2v/u$. Differentiating (\ref{fnpd2})
with respect to $v$, we then obtain
\begin{align}
&2^{-3\alpha}\left(
-\frac{2v}{u}{\>}
g_{\alpha}^{\prime}(u){\,}g_{\alpha}(v)^{2}
+2g_{\alpha}(u){\,}g_{\alpha}(v){\,}g_{\alpha}^{\prime}(v)
\right)
\nonumber\\
&=2\alpha{v}\wff_{\alpha}(u,v)
\left(
\frac{f_{\alpha}(u)}{g_{\alpha}(u)}-
\frac{f_{\alpha}(v)}{g_{\alpha}(v)}
\right)
{\,}. \label{fnpd22}
\end{align}
We used here that
$u^{-1}g_{\alpha}^{\prime}(u)=-\alpha{f}_{\alpha}(u)$ due to
(\ref{fudf}) and (\ref{gudf}). As was already shown, the ratio
$f_{\alpha}(u)/g_{\alpha}(u)$ is monotonically increasing function
of $u\in[0;1]$. So, the derivative (\ref{fnpd22}) vanishes only
for $u=v$. Together with (\ref{uvcd}), we get $u=v=1/\sqrt{3}$.
For $v<1/\sqrt{3}<u$, the derivative (\ref{fnpd22}) is strictly
positive; for $u<1/\sqrt{3}<v$, the derivative (\ref{fnpd22}) is
strictly negative. At the point $u=v=1/\sqrt{3}$, therefore, the
function (\ref{fnpd2}) reaches its conditional maximum
\begin{equation}
\max\wff_{\alpha}=\left(2^{-\alpha}
g_{\alpha}\bigl(1/\sqrt{3}\bigr)\right)^{3}
{\>}. \label{cmwf}
\end{equation}
Combining this with (\ref{fnpd1}) completes the proof. $\blacksquare$

\begin{figure}
\includegraphics{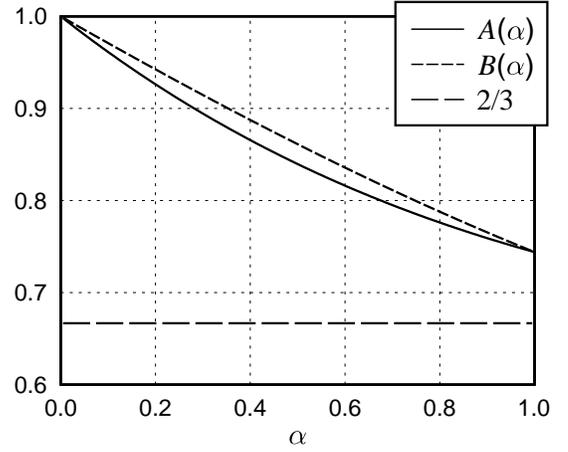}
\caption{The upper bounds $A(\alpha)$ and $B(\alpha)$ as functions
of $\alpha$.} \label{fig001}
\end{figure}

In the case of pure measured state, the statement of Theorem
\ref{upral} gives tight upper bounds on the entropic sum for all
real $\alpha\in(0;1]$. Note that these bounds can be motivated by
some plausible reasons. Let us consider the entropic value, which
is averaged over the individual ones. It is also useful to
rescale each entropy by its maximal possible value $\ln{2}$.
Combining (\ref{ral0}) and (\ref{urwh}), we finally obtain
\begin{equation}
\frac{2}{3}\leq
\frac{1}{3\ln{2}}\sum_{\nu=x,y,z} R_{\alpha}\bigl(\bsg_{\nu}|\psi\bigr)
\leq{B}(\alpha)=\frac{\wrh_{\alpha}}{\ln{2}}
\ . \label{lour12}
\end{equation}
These bounds hold for $\alpha\in(0;1]$ and arbitrary pure state
$|\psi\rangle$ of qubit. The lower and upper bounds of
(\ref{lour12}) are both tight in the sense that they are reached
under the certain conditions for equality. In the paper
\cite{rastqip13}, we have derived the bounds with Tsallis'
entropies. Denoting
\begin{equation}
\wth_{\alpha}=\frac{1}{1-\alpha}{\,}\left\{
\left(\frac{1+1/\sqrt{3}}{2}\right)^{\alpha}
+\left(\frac{1-1/\sqrt{3}}{2}\right)^{\alpha}-1
\right\}
{\,}, \label{prone}
\end{equation}
for real $\alpha\in(0;1]$ and integer $\alpha\geq2$ we have
\begin{equation}
\frac{2}{3}\leq
\frac{1}{3\ln_{\alpha}(2)}\sum_{\nu=x,y,z} H_{\alpha}\bigl(\bsg_{\nu}|\psi\bigr)
\leq{A}(\alpha)=\frac{\wth_{\alpha}}{\ln_{\alpha}(2)}
\ . \label{loup12}
\end{equation}
In both the Tsallis and R\'{e}nyi cases, the lower bound on the
rescaled average $\alpha$-entropy is equal to $2/3$. It is
instructive to compare the corresponding upper bounds $A(\alpha)$
and $B(\alpha)$. On Fig. \ref{fig001}, these bounds are shown as
functions of $\alpha\in(0;1]$. The upper bounds $A(\alpha)$ and
$B(\alpha)$ are very close to each other, though
$A(\alpha)<B(\alpha)$ for all $\alpha\in(0;1)$. The difference
between them does not exceed 2.5{\%} in a relative scale. Although
the value $\alpha=0$ itself is not considered, we have
$A(\alpha)\to{1}^{-}$ and $B(\alpha)\to{1}^{-}$ in the limit
$\alpha\to{0}^{+}$. In the case $\alpha=1$, we
obtain $A(1)=B(1)\approx0.744$. At the constant lower
bound, the upper bounds $A(\alpha)$ and $B(\alpha)$ monotonically
decreases with $\alpha$. So, the bands are reducing with growth
of $\alpha$. A width of the corresponding band may be interpreted
as a measure of sensitivity in quantifying the complementarity.
From this viewpoint, there is no significant distinction between
the Tsallis and R\'{e}nyi formulations for $\alpha\in(0;1)$. In
particular problems, we often have specific reasons for choosing
an appropriate entropic measure.

\section{Conclusions}\label{sec5}

We have examined uncertainty and certainty relations for the Pauli
observables in terms of the R\'{e}nyi $\alpha$-entropies of order
$\alpha\in(0;1]$. These relations are respectively formulated as
lower and upper bounds on the sum of three $\alpha$-entropies. The
bounds are tight in the sense that they can certainly be
saturated. Explicit conditions for equality are obtained as well.
As the Renyi $\alpha$-entropy is a non-increasing function of
$\alpha$, the lower bounds for $\alpha\in(0;1]$ could be derived
from the previous results on the Shannon entropies. However,
conditions for equality cannot be obtained in this way. In the
case of pure measured states, tight upper bounds on the sum of
three R\'{e}nyi's entropies are further derived. We also
discussed the interval, in which the rescaled average R\'{e}nyi
entropy of order $\alpha\in(0;1]$ ranges in the pure-state case.
This interval has been compared with the Tsallis formulation.

\end{document}